\documentclass[prl,aps,amssymb,amsfonts,amsmath,doublecolumn,superscriptaddress,preprint]{revtex4-1} 
\usepackage{graphicx} 
\usepackage{amsmath} 
\usepackage{amsfonts}

\newcommand{\mplaffil}{Max Planck Institute for the Science of Light, 91058 Erlangen, Germany}
\newcommand{\DMAVTaffil}{Laboratory of Thermodynamics in Emerging Technologies, ETH Zurich, 8092 Zurich, Switzerland}
\newcommand{\ethaffil}{Laboratory of Physical Chemistry, ETH Zurich, 8093 Zurich, Switzerland}
\newcommand{\FAUaffil}{Department of Physics, Friedrich-Alexander University Erlangen-N\"urnberg, 91058 Erlangen, Germany}
\newcommand{\KGnew} {Department of Physics, Hanyang University, 222 Wangsimni-ro, Seongdong-gu, Seoul, 133-791, Korea}

\begin{document} 
\title{Spontaneous emission enhancement of a single molecule by a double-sphere nanoantenna across an interface} 
\author{K-G. Lee*}
\affiliation{\mplaffil}\affiliation{\KGnew}
\author{H. Eghlidi*}
\affiliation{\DMAVTaffil}
\author{X-W. Chen}
\affiliation{\mplaffil}
\author{A. Renn}
\affiliation{\ethaffil}
\author{S. G\"otzinger}
\affiliation{\FAUaffil}\affiliation{\mplaffil}
\author{V. Sandoghdar}
\affiliation{\mplaffil}\affiliation{\FAUaffil} \email{vahid.sandoghdar@mpl.mpg.de}

\begin{abstract} 
We report on two orders of magnitude reduction in the fluorescence lifetime when a single molecule placed in a thin film is surrounded by two gold nanospheres across the film interface. By attaching one of the gold particles to the end of a glass fiber tip, we could control the modification of the molecular fluorescence at will. We find a good agreement between our experimental data and the outcome of numerical calculations. 
\end{abstract}

\maketitle

Spontaneous emission plays a central role in the majority of optical phenomena in our world. The rate of this process is determined by the radiative decay of the excited state of the emitter and scales with the second and third powers of its dipole moment and transition frequency, respectively \cite{Loudon}. Because optical emitters have nanometer or subnanometer extensions, the size of their dipole moments is restricted, giving rise to typical fluorescence lifetimes in the nanosecond range, which is orders of magnitude longer than an optical cycle. This long lifetime and the correspondingly slow decay restrict the optical power that can be extracted from a single emitter. To get around this limitation, one can devise schemes for enhancing the spontaneous emission rate.

After the seminal work of Drexhage more than forty years ago \cite{drexhage74}, a flurry of theoretical and experimental works investigated the modification of the spontaneous emission rate in microcavities \cite{Berman-book,Yokoyama-book,chang96}. The challenge in this approach in achieving very large effects translates to the realization of cavities with very high quality factor $Q$ and low mode volume $V$. Although optical microcavities have been used in a variety of contexts, their performance is restricted by two fundamental issues. First, they are at least several micrometers in size because highly reflective mirrors take some space. Second, high $Q$s imply narrow resonances, dictating operation with very narrow-band emitters and the need for highly sensitive frequency tuning. To our knowledge, the largest cavity-assisted enhancement reported so far amounts to about 60 \cite{press07}. 

Optical nanoantennas provide ideal alternatives for very strong enhancement of spontaneous emission with subwavelength footprints and broadband response \cite{Antenna-book-12}. The main experimental challenges in this system are the needs for 1) careful and high-quality fabrication of antenna nanostructures, 2) nanometer precision in positioning the emitter and 3) alignment of its dipole moment with respect to the nanoantenna. About a decade ago, we addressed these difficulties by introducing a scannable nanoantenna, consisting of a single plasmonic nanosphere attached to the end of a dielectric tip \cite{Kalkbrenner:01, Kalkbrenner:05}. The simple and symmetric form of the spherical nano-antenna and the \textit{in-situ} nanometric position control provided by scanning probe technology have opened the door to quantitative measurements and comparison with theory \cite{Anger:06, Kuehn:06, eghlidi09, Bharadwaj:10}. 

Although a simple gold nanoparticle provides a convenient and powerful model system for optical antennas, it does not result in very large enhancement effects. Employment of antennas made of two nanostructure components, on the other hand, promises much larger modifications of up to three orders of magnitude \cite{Rogobete:07}. Following the virtues of a scanning probe approach, scientists have placed double-particle antennas at the end of a tip \cite{Farahani:05} and assembled plasmonic nanoparticles into desirable positions surrounding an emitter on a surface \cite{schietinger09}. However, previous efforts confronted the problem that  the finite size of the two antenna components prevents one from overlapping their hot spot with an emitter placed on the sample surface \cite{Farahani:05}. As a result, only enhancements up to about ten times have been reported in this fashion. An alternative strategy has been to spread molecules on arrays of pre-fabricated antennas \cite{kinkhabwala09}, followed by a search for very rare events, where the position and orientation of an emitter suit those of an antenna. In addition, this approach does not allow one to vary any of the key parameters for control experiments. In this Letter, we combine various features of the above-mentioned techniques to perform quantitative measurements on single molecules sandwiched between two spherical gold nanoparticles. 

\begin{figure*}
\begin{center}
\includegraphics[scale=0.8]{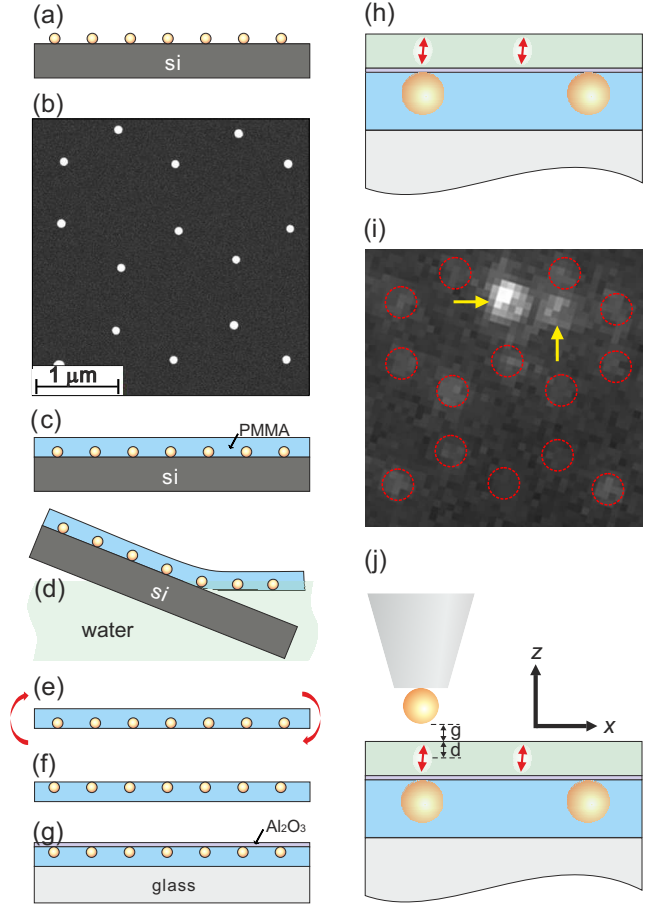}
\caption{Various steps for the preparation of an array of gold nanoparticles embedded just under a dielectric surface. a) An array of gold nanoparticles fabricated via thermal annealing of a triangular pattern made by colloidal lithography and gold deposition. b) scanning electron microscope image of a section of the substrate sketched in (a). c) Coating with PMMA. d) Floating of the film on water. e, f) Flipping. g) Atomic Layer Deposition (ALD) of a thin aluminum oxide layer. h) Arrangement of the sample after spin coating a thin film of pT containing terrylene molecules. i) Fluorescence image showing two terrylene molecules; one of them lies on top of a gold nanoparticle and is enhanced. j) Arrangement of the setup including the scanning tip with a gold nanoparticle. } \label{scheme}
\end{center}
\end{figure*}

One of the major difficulties in our work has been sample preparation. Figure \ref{scheme} shows the schematics of the experimental procedure. We start by colloidal self-assembly on a silicon wafer followed by evaporation of 30 nm of gold to obtain a hexagonal lattice of gold islands \cite{Fischer:81}. After removing the polymer colloids, we anneal the sample at 1100 ¡C for  about 40 minutes to convert the triangular metallic islands to quasi-spherical features, whereby the diameter of the latter (110 nm) is controlled by the thickness of the initial gold deposition and annealing duration \cite{Tan:05}. Next, we cover the gold nanoparticles with PMMA, float the resulting film in water, flip it and place it on a glass substrate \cite{Kalkbrenner:00}. This yields a very flat film, in which the embedded gold nanoparticles touch the upper surface. The next challenge is to spincoat a thin layer of p-terphenyl (pT) as a crystalline host for fluorescent terrylene molecules \cite{Pfab:04}. To protect the PMMA layer from dissolving during this step, we cover it with about 4 nm of $\rm{Al}_2\rm{O}_3$ via atomic layer deposition. 

The resulting structure consists of a hexagonal array of gold nanospheres (see Fig. \ref{scheme}g) covered with a pT film of about 25 nm, containing a low concentration of single terrylene molecules. Figure \ref{scheme}i illustrates a fluorescence image.  In this example, one single molecule happens to be located on top of one of the gold nanoparticles (marked by the horizontal arrow) while another lies away from gold nanospheres (marked by the vertical arrow). To record this fluorescence image, the sample was illuminated in total internal reflection through an oil-immersion microscope objective at the wavelength of 532 nm. Terrylene emission was collected via the same objective and detected on a CCD camera or with an avalanche photodiode (APD). By using a pulsed laser with pulse widths of about 10 ps, we could record the fluorescence lifetime of the excited state. We remark that even the gold nanoparticles can be seen in Fig. \ref{scheme}i because of their faint fluorescence \cite{Geddes:03}. 

\begin{figure*}
\begin{center}
\includegraphics[scale=0.8]{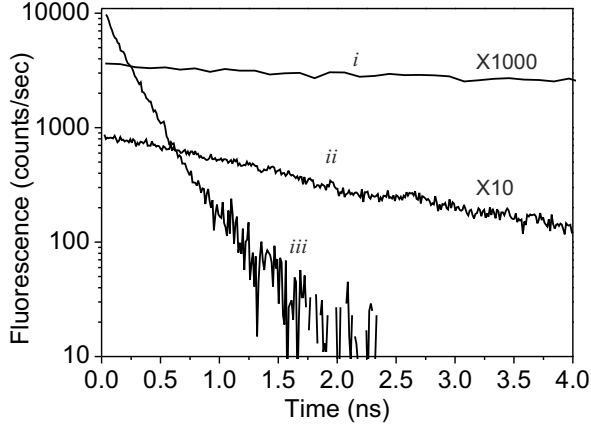}
\caption{The fluorescence lifetime of a single molecule in a thin p-terpheyl film (i) in the absence of gold nanoparticles, (ii) on top of a single gold nanoparticle, (iii) sandwiched between two nanoparticles (see Fig. \ref{scheme}j).} \label{data}
\end{center}
\end{figure*}

The fluorescence lifetime ($\tau$) of terrylene in bulk pT is about 4 ns \cite{harms99}. However, the lifetime is lengthened to $20 \pm 5$ ns for molecules in thin pT films, depending on the exact depth of the molecule (see curve \textit{i} in Fig. \ref{data}). This is because the radiative decay rate of a dipole perpendicular and close to an interface is lowered if the dipole is placed in the high-index side \cite{Lukosz1977}. In the language of the local density of states (LDOS), this is equivalent to a lowering of LDOS. 

In case of terrylene in pT, the transition dipole moment is tilted by about 15 degrees \cite{Pfab:04}. This out-of-plane alignment of the terrylene dipole moment means that it is also oriented radially with respect to a gold nanoparticle situated under the molecule. In this configuration, the fluorescence lifetime is shortened \cite{Ruppin:82}. Curve \textit{ii} in Fig. \ref{data} shows a decay curve for a single molecule in such a situation, reaching a lifetime of $\tau=1.5$ ns. This enhanced decay is accompanied by an increase of 3 times in the fluorescence of the molecule as compared to molecules not coupled to plasmonic structures. It is this small enhancement together with the coinciding diffraction-limited images of a terrylene molecule and a gold nanoparticle that let us identify cases, where they lie on top of each other (see Fig. \ref{scheme}i). We note in passing that the observed enhancement is weaker than our previous studies \cite{Kuehn:06, eghlidi09} by an order of magnitude because here the nanoparticle is also inside the pT film, where the LDOS is diminished, leading to weaker scattering and radiation \cite{eghlidi09, Chen:12}.

Once we identified a case of coupling between a molecule and a nanoparticle in the substrate, we introduced a second gold nanoparticle attached to the end of a glass fiber tip \cite{Kalkbrenner:01}, as illustrated in Fig. \ref{scheme}j. Curve \textit{iii} in Fig. \ref{data} plots the resulting fluorescence decay with a $1/e$ time of $\tau=180$ ps, which corresponds to a 111-fold reduction of the fluorescence lifetime as compared to 20 ns for an unperturbed molecule. Here, we accounted for the finite time resolution of 60 ps caused by the response time of the APD and the laser pulse. To do this, we impinged laser pulses on the sample substrate and evaluated their reflection by the bare surface. The shortening of the lifetime is now accompanied by a fluorescence enhancement. Our steady-state measurements of the fluorescence signal yielded enhancement factors of 3 and 28 for the cases of bottom particle only and both particles, respectively. These changes also manifest themselves in the y-intercepts of the decay curves in Fig. \ref{data} because for a simple exponential process the area under the decay curve represents the radiated energy and is, thus, conserved.

\begin{figure*}
\begin{center}
\includegraphics[scale=0.8]{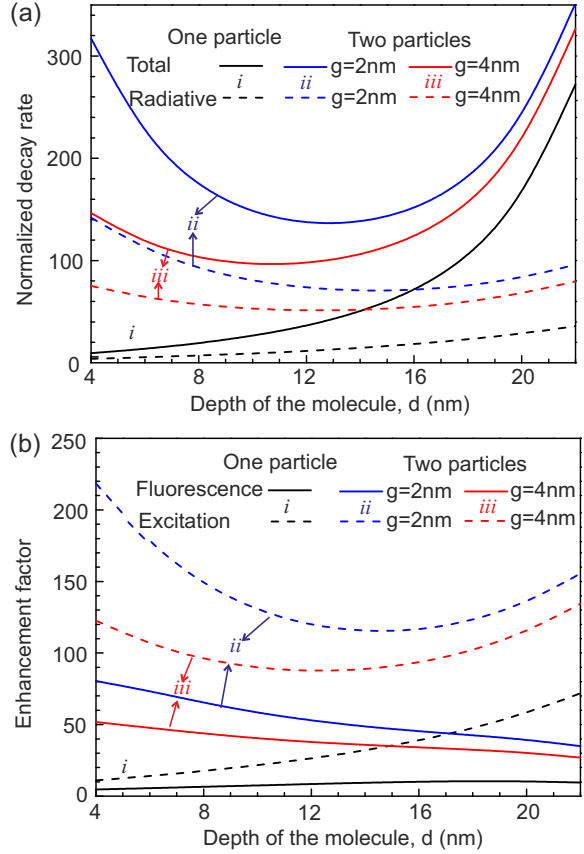}
\caption{Numerical calculations (BOR-FDTD) of the decay rates and the enhancement factors of excitation and fluorescence for different configurations (see also Fig \ref{scheme}j). The pT film thickness was taken to be 25 nm. See legends for details.} \label{theory}
\end{center}
\end{figure*}

As we recently pointed out, the presence of a dielectric interface can considerably affect the interaction between the emitter and the plasmonic antenna \cite{eghlidi09,Chen:12}. To take this and the effect of the tip shaft into account, we have performed rigorous body of revolution finite-difference time-domain (BOR-FDTD) calculations, whereby we take advantage of the cylindrical symmetry of the problem to gain efficiency. The black solid and dashed curves labeled \textit{i} in Fig. \ref{theory}a display the changes of the total fluorescence decay rate and the radiative decay rate, respectively, in the absence of the upper nanoparticle. As expected, the decay rate is enhanced if the separation between the molecule and the lower nanoparticle is decreased, i.e. the depth $d$ of the molecule inside the pT film is increased (see Fig. \ref{scheme}j). The blue and red curves labeled \textit{ii} and \textit{iii}, respectively, show the corresponding quantities for two different positions of the upper gold nanoparticle. The shortening of the lifetime for large $d$ values is caused by the dominating nonradiative decay rate when the particle becomes close to the lower nanoparticle.

The fluorescence signal $S$ of an emitter can be written as $S \propto \xi.K.\eta.I$ in the weak excitation regime. Here $\xi$ represents the collection efficiency, $I$ is a measure for the excitation intensity, $K$ stands for the enhancement of the excitation intensity at the emitter position and along its dipole moment, and $\eta$ is the quantum efficiency. Our calculations confirm that $\xi$ is reduced from $80\%$ to $60\%$ by adding the top particle and the glass fiber tip shaft. This change is small because the presence of a spherical nanoparticle in the near field of the molecule mostly maintains its dipolar radiation characteristics. The dashed and solid curves in Fig. \ref{theory}b plot the excitation enhancement and total fluorescence enhancement for the same conditions as in Fig. \ref{theory}a, respectively. 

We can now try to find the conditions which simultaneously match the experimental values of $S_{\rm{0}}$, $S_{\rm{1}}$, $S_{\rm{2}}$, $\tau_{\rm{0}}$, $\tau_{\rm{1}}$, and $\tau_{\rm{2}}$, where the indices 0, 1 and 2 refer to the cases of no, one and two gold nanoparticles, respectively. To account for relative lateral displacement of the two particles and the molecule (BOR-FDTD calculations only treat geometries with axial symmetry) we performed full three-dimensional FDTD calculations to allow for a displacement of the molecule with respect to the lower nanoparticle in the substrate. The position of the top particle is controlled in the experiment and can be optimized. By calculating the decay rates for dipole moments parallel and perpendicular to the substrate, we also accounted for the $15^{\circ}$ tilt of the dipole moment of terrylene in pT. 

We find that the depth of the molecule $d=6$ nm, the gap between the upper particle and the pT interface $g=4$ nm, and a lateral offset of 12 nm between the molecule and the axis of the bottom particle provide $S_{\rm{1}}/S_{\rm{0}}=4.4$, $S_{\rm{2}}/S_{\rm{0}}=40$, $\tau_{\rm{1}}/\tau_{\rm{0}}=11$, and $\tau_{\rm{2}}/\tau_{\rm{0}}=114$, which match the experimental values of $S_{\rm{1}}/S_{\rm{0}}=3$, $S_{\rm{2}}/S_{\rm{0}}=28$, $\tau_{\rm{1}}/\tau_{\rm{0}}=11.5$, and $\tau_{\rm{2}}/\tau_{\rm{0}}=111$ well. The tolerance in obtaining this match is very tight. Nevertheless, the deduced values of $d$ and $g$ do not offer a unique solution because the exact thickness of the pT film might be different from the assumed value of 25 nm. Furthermore, the deviation of the shape of the nanoparticles embedded in the substrate from a sphere can affect the exact values. However, we did confirm that these nanoparticles had plamson spectra comparable with those of colloidal spheres of the same size. 

The quantum efficiency of terrylene molecules in bulk pT crystals is close to one \cite{Buchler:05}. However, the lengthened lifetime of molecules very close to an interface reduces the quantum efficiency to values between $80\%$ and $90\%$, depending on the exact depth of the molecule. When the lower particle is added, calculations indicate that the quantum efficiency of the molecule studied above was reduced further to $38\%$. This value was then increased to $54\%$ when the top particle was approached. This corresponds to an overall enhancement factor of 64 for the radiative decay (spontaneous emission) rate. We note that quantitative experimental separation of the radiative and nonradiative decay channels is a challenging task because it requires an accurate and independent measurements of the collection efficiency $\xi$ and excitation enhancement $K$. 

\begin{figure*}
\begin{center}
\includegraphics[scale=0.5]{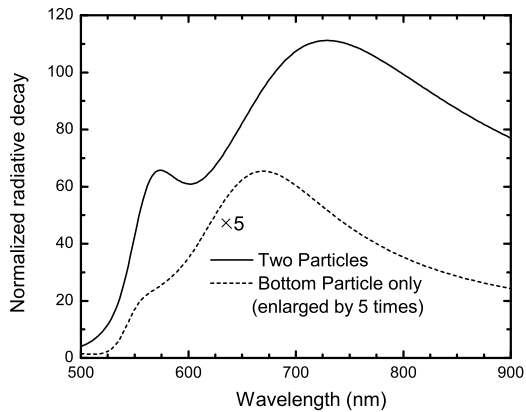}
\caption{Calculated enhancement of the radiative decay rate as a function of wavelength for a single molecule in the pT film in the presence of the lower antenna particle (dashed) and for both lower and top particles (solid).} \label{plasmon}
\end{center}
\end{figure*}

Another issue to be considered in our experiment is the role of the plasmon spectrum of the antenna. The plasmon spectrum of an isolated gold nanoparticle with a diameter of 50-100 nm is known to lie in the region of 530 nm, but the center wavelength and width of the spectrum change in a very sensitive fashion according to the particle environment. As a result, the spectra of the lower and upper antenna particles across the dielectric interface are expected to be different. In addition, measured plasmon spectra of finite-sized objects depend on the way they are excited \cite{mojarad09a}. To explore the effect of the plasmon spectra ``seen" by a terrylene molecule sandwiched in our antenna geometry, in Fig. \ref{plasmon} we plot the results of BOR-FDTD calculations for the variation of the radiative decay rate with wavelength. We see that the combination of our antenna structure and molecule is not optimized; an emitter with transition around 750 nm would have experienced about twice higher enhancement of the spontaneous emission. Another option would have been to use silver as the material of choice to tune the resonance more towards the lower wavelength.

In conclusion, we have presented experimental and theoretical results on the coupling of a single molecule to a double-particle plasmonic antenna. While previous attempts had studied arrangements, where antenna components lied on a substrate, we have examined the case where a dielectric interface cuts through the antenna. By controlling the position of one of the two particles, we could obtain sufficient data to compare our experimental measurements with theory, thus, deducing a 64-fold enhancement of the spontaneous emission rate and a final quantum efficiency of $54\%$. We emphasize that these measurements were performed on a molecular system with a near-unity intrinsic quantum efficiency \cite{Buchler:05}. The antenna effect would have been more dramatic for emitters with lower quantum efficiency \cite{kinkhabwala09}. For example, the same enhancement factor would have improved an intrinsic quantum efficiency of $1\%$ by 43 folds \cite{agio07, Bharadwaj:10}. The combination of such an improvement with an on-command activation through mechanical actuation of the antenna has interesting prospects in photonics.

K-G Lee and H. Eghlidi contributed equally to this work.

\bibliography{vahid} 
\end{document}